\documentstyle[amssymb,preprint,aps]{revtex}
\begin{document}
\draft
\title{Experimental Implementation of Hogg's Algorithm on a Three-Quantum-bit NMR
Quantum Computer}
\author{Xinhua Peng$^{1}$, Xiwen Zhu$^{1\thanks{%
Corresponding author. {E-mail:xwzhu@nmr.whcnc.ac.cn; Fax: 0086-27-87885291.}}%
}$, Ximing Fang$^{1,2}$, Mang Feng$^{1}$, Maili Liu$^{1},$ and Kelin Gao$%
^{1} $}
\address{$^{1}$Laboratory of Magnetic Resonance and Molecular Physics, Wuhan
Institute of Physics and Mathematics, The Chinese Academy of Sciences,
Wuhan, 430071, People's Republic of China\\
$^{2}$Department of Physics, Hunan Normal University, Changsha, 410081,
People's Republic of China}
\maketitle

\begin{abstract}
Using nuclear magnetic resonance (NMR) techniques with three-qubit sample,
we have experimentally implemented the highly structured algorithm for the
1-SAT problem proposed by Hogg. A simplified temporal averaging procedure
was employed to the three-qubit spin pseudo-pure state. The algorithm was
completed with only a single evaluation of structure of the problem and the
solutions were found with probability 100$\%$, which outperform both
unstructured quantum and the best classical search algorithm.
\end{abstract}


\vskip 1cm

\narrowtext

\section{\bf Introduction}

Quantum Computers exploiting the principles of quantum parallelism and
entanglement of quantum states outperform classical counterparts greatly.
Quantum algorithms can solve computational problems much faster than the
classical ones\cite{Deutsch,Cleve,Shor,Grover}, and can solve even some
intractable problems to classical methods, such as the most famous problem
of factoring large integers\cite{Shor}. Hitherto, the selected systems to
realize quantum computation have been proposed such as trapped ions\cite
{Cirac,Monroe}, quantum dots\cite{Bandyopadhyay}, and cavity
quantum-electrodynamics\cite{Domokos,Turchette}. However, the most effective
one to realize quantum algorithms to date is the liquid-state NMR spins\cite
{Cory1,Gershenfeld}. Several quantum algorithms have been implemented by
liquid-state NMR ensemble including the Deutsch-Jozsa algorithm with two,
three\cite{ChuangDJ,Jones1,Linden,Kim,Collins} and five\cite{Marx} qubits,
the Grover's algorithm with two\cite{Chuang3,Jones2} and three qubits\cite
{L.M.K}, an algorithmic benchmark\cite{Knill1}, an order-finding algorithm
with five qubits\cite{Lieven}.

Search algorithm is one of the most important algorithms. The difference
between quantum and classical search algorithms is that, quantum ones can
operate simultaneously on a superposition of all possible search states so
as to be superior to classical counterparts greatly. A quantum search
algorithm that uses the underlying structure of the search problem to
establish correlations between problem properties and the solutions results
in a substantial improvement over a search one that ignores them, such as
Hogg's algorithm\cite{Hogg1} vs. Grover's one\cite{Grover}. However, when
only two qubits are considered, there is no difference between Hogg's
algorithm and Grover's one because both of the algorithms only require one
search step. For $n>2$ qubits, the superiority of Hogg's algorithm will be
displayed, which still require a single step, while Grover's one requires $%
\sqrt{2^{n}}$ steps. For example, when $n=3$ qubits, Grover's algorithm
needs at least $2$ steps with probability near $95\%$. Hogg's algorithm for
2-qubit has been implemented in Ref. \cite{Zhu}. In this paper, we
experimentally implemented the Hogg's algorithm for a 1-SAT with the 3-qubit
pseudo-pure state prepared by three-step temporal averaging.

\section{The highly structured algorithm}

The highly structured algorithm is associated with the satisfiability
problem (SAT)\cite{Garey}, one of the most difficult class of
non-deterministic polynomial problem (NP) \cite{Garey}. A SAT is a
combinatorial search problem\cite{Garey}, consisting of a logic formula in $%
n $ boolean variables (true or false) $V_{1,}...,V_{n}$ and the requirement
to find an assignment, specifying a value for each boolean variable, that
makes the formula true. The logic formula can be expressed as a conjunction
of $m$ clauses and each clause is a disjunction of some variables, any of
which may be negated. When all the clauses have $k$ variables, the problem
is called $k $-SAT. In general, the computational cost of solving a SAT
grows exponentially with $n$ in the worst case. For $k\geqslant 3,$ $k$-SAT
is NP-complete. However, for 1-SAT and maximally constrained k-SAT, Hogg
proposed a quantum algorithm to find a solution in one step by using problem
structure. As each false clause for a given assignment is counted as a
conflict, solutions are assignments with no conflict. Therefore, the
correlations between problem properties and the assignments are set up so
that Hogg's algorithm can solve problems with only a single testing of all
the assignments, corresponding to a single classical search step.

For a 1-SAT problem with $n$ variables and $m$ clauses, in fact, the
algorithm takes the following four stages\cite{Hogg1}.

1. Initialize the amplitude equally among the states, giving an equal
superposition of bases $|s\rangle $, $|\psi _{i}\rangle =2^{-\frac{n}{2}%
}\sum_{s}|s\rangle $ , with bit strings s being all $2^{n}$ assignments of $%
n $ variables.

2. Adjust the phases of $|s\rangle $ based on the conflicts $c$ in the
assignments $s$, ranging from $0$ to $m$, i.e., apply a transformation $R$
on $|\psi _{i}\rangle ,$ where $R$ is a diagonal matrix with elements 
\begin{equation}
R_{ss}=\left\{ 
\begin{array}{cc}
\sqrt{2}\cos [(2c-1)\frac{\pi }{4}], & \text{ }for\text{ }even\text{ }m; \\ 
i^{c}, & for\text{ }odd\text{ }m.
\end{array}
\right.
\end{equation}

3. Mix the amplitudes from different assignments with the mixing matrix $U$
depending only on the Hamming distance $d$ between $r$ and $s$ is described
as

\begin{equation}
U_{rs}=U_{d(r,s)}=\left\{ 
\begin{array}{cc}
2^{-\frac{n-1}{2}}\cos [(n-m+1-2d)\frac{\pi }{4}], & for\text{ }even\text{ }%
m; \\ 
2^{-\frac{n}{2}}e^{i\pi (n-m)/4}(-i)^{d}, & for\text{ }odd\text{ }m.
\end{array}
\right.
\end{equation}
Here, $d(r,s)=\left| r\right| +\left| s\right| -2\left| r\wedge s\right| $
with $\left| r\right| $ $\left( \left| s\right| \right) $ being the number
of 1-bits in $r(s)$ and $\left| r\wedge s\right| ,$ the number of 1-bits
both assignments have in common. The operator $U$ can be defined in terms of
two simpler operations: $U=W\Gamma W,$ where $W=2^{-\frac{n}{2}}(-1)^{\left|
r\wedge s\right| }$ is the Walsh-Hadamard transform, $,$ and $\Gamma $ is a
diagonal matrix with elements 
\begin{equation}
\Gamma _{rr}=\gamma (r)=\gamma _{h}=\left\{ 
\begin{array}{cc}
\sqrt{2}\cos [(m-2h-1)\frac{\pi }{4}], & \text{ }for\text{ }even\text{ }m;
\\ 
i^{h}e^{-i\pi m/4}, & for\text{ }odd\text{ }m.
\end{array}
\right.
\end{equation}
depending only on the number of 1-bits in each assignments $h=\left|
r\right| ,$ ranging from $0$ to $m$.

4. Measure the final superposition $|\psi _{f}\rangle =UR|\psi _{i}\rangle .$

For our 3-spin system (i. e. $n=3$), the number of clauses , $m$, in the
1-SAT formula can be 3, 2, or 1. The quantum circuit for the highly
structure search in a 3-qubit system is shown in Fig. 1. According to four
stages of the Hogg's algorithm above, the calculation starts with a
Walsh-Hadamard transform $W$ applying to a pseudo-pure state $|000\rangle $,
to prepare the equal superposition state $|\psi _{i}\rangle ,$ where $%
W=H_{A}\otimes H_{B}\otimes H_{C},$with $H=X^{2}Y$ (pulses applied from
right to left) being a single-spin Hadamard gate. Then we derived the NMR
pulse sequences for $R$ and $U$ in Eqs .(1) and (2). We take the 1-SAT with $%
m=3 $, i.e., the logic formula being $V_{1}\wedge V_{2}\wedge V_{3}$, as an
example. In this case, $U$ could be represented by $W\Gamma W.$ From Eqs.(1)
and (3), the expressions for $R$ and $\Gamma $ can be written into

\begin{equation}
diag\left[ R_{V_{1}\wedge V_{2}\wedge V_{3}}\right] =[-i,-1,-1,i,-1,i,i,1],
\end{equation}

\begin{equation}
diag\left[ \Gamma _{m=3}\right] =[1,i,i,-1,i,-1,-1,-i].
\end{equation}
which correspond to the pulse sequences $R_{V_{1}\wedge V_{2}\wedge
V_{3}}\backsim \bar{Z}_{1}\bar{Z}_{2}\bar{Z}_{3}$, $\Gamma _{m=3}\backsim
Z_{1}Z_{2}Z_{3}$ (up to an irrelevant overall phase factor). In order to
make the best use of the available coherence time and to diminish errors due
to the increased number of RF pulses, the pulse sequences of $W,$ $R$ and $%
\Gamma $ were optimized to eliminate unnecessary operations with the help of
NMR\ principle \cite{Ernst}. Two instances for $m=1$ and $3$ were chosen for
demonstrating Hogg's algorithm in our experiments for their simple pulse

sequences. All the reduced pulse sequences for all possible formulas with $%
m=1 $ and $3$ as well as theoretical solutions are listed in Table. 1.

\section{The NMR experimental results}

The Hogg's algorithm was implemented by liquid-state high-resolution NMR
spectroscopy with carbon-13 labeled alanine $NH_{3}^{+}-C^{\alpha
}H(C^{\beta }H_{2})-C^{^{\prime }}O_{2}^{-}$ dissolved in $D_{2}O$. We chose 
$C^{^{\prime }}$ , $C^{\alpha }$ and $C^{\beta }$ as the three-spin system
in experiments, representing spin 1, spin 2 and spin 3, respectively.
Spectra were recorded on a Bruker ARX500 spectrometer with a probe tuned at
125.77MHz for $^{13}C.$ The chemical shifts of three different carbon spins
were calibrated about $-4320,$ $0,$ and $15793$Hz, and the spin-spin
coupling constants $J_{12},$ $J_{23},$ and $J_{13}$, 34.94, 53.81, and
1.21Hz, respectively. The relaxation times were measured to be $%
T_{1}(1)=20.3\sec ,$ $T_{1}(2)=2.8\sec ,$ and $T_{1}(3)=1.5\sec $ and $%
T_{2}(1)=1.3\sec $, $T_{2}(2)=0.41\sec $, and $T_{2}(1)=0.81\sec $\cite
{Collins}. Protons were decoupled during the whole experiments, using a
standard heteronuclear decoupling technique. Spin-selective excitations were
executed using low-power, long-duration pulses with a {\it Gaussian} shape.
The length of these pulses was tailored to achieve sufficient selectivity in
the frequency domain without disturbing the nearest nucleus, depending on
the difference of the chemical shifts between nuclei. We took the length of
the selective pulses to be 0.7ms and the excitation power to be 29.6dB for
selective $\frac{\pi }{2}$ pulses for all $^{13}C$ nuclei.

\subsection{The preparation of pseudo-pure states}

In liquid-state NMR ensemble quantum computers, Chuang et al.\cite
{Gershenfeld} proposed pseudo-pure states instead of the genuine pure states
as the initial state. Several methods have been proposed to prepare the
pseudo-pure states including spatial averaging \cite{Cory1}, temporal
averaging \cite{Knill}, logical labeling\cite{Gershenfeld,Chuang4,Dorai}.
Recently, an alternative simplified temporal averaging was proposed by
Lieven et al.\cite{Lieven}. For a $n$-qubits homonuclear system, the
deviation density matrix of the system in thermal equilibrium is a sum of $n$
terms, 
\begin{equation}
\rho _{eq}=\sum\limits_{k=1}^{n}I_{kz}.
\end{equation}
The deviation density matrix of the desired pseudo-pure state $%
|00...0\rangle $ is

\begin{equation}
\rho
_{eff}=\sum\limits_{k=1}^{n}I_{kz}+\sum\limits_{l>k=1}^{n}I_{kz}I_{lz}+\sum%
\limits_{m>l>k=1}^{n}I_{kz}I_{lz}I_{mz}+......,
\end{equation}
a sum of $2^{n}-1$ terms. In order to obtain the desired pseudo-pure state, $%
[(2^{n}-1)/n]$ different experiments are at least needed by utilizing
controlled-NOT ($CN_{ij}$ flips spin $j$ if and only if $i$ is $|1\rangle $
) and NOT ($N_{i}$ flips the sign of spin $i$).

For a homonuclear 3-spin system, (i.e. $n=3$), three separate experiments
are at least required to achieve the pseudo-pure state $\rho _{000}$. We use
3 separate experiments, giving a total of 9 product operator terms. The 2
extra of 9 terms are canceled out pairwise by using NOT $\left( N_{i}\right) 
$ operations. The procedure can be chosen in a variety of ways. For
different samples, one can choose an optimal experimental scheme. As to our
sample of alanine with $1/2J_{13}=0.41\sec ,$ which is comparable to the
smallest $T_{2},$ gate $CN_{13}$ or $CN_{31}$ containing the evolution under
the scalar coupling $J_{13}$ should be excluded in the preparation circuit.
We choose the following 3 states preparation sequences acting on the
equilibrium as

\begin{equation}
E,\qquad CN_{32}CN_{21}N_{3},\qquad CN_{21}CN_{12}CN_{32}
\end{equation}
where $E$ is unit matrix denoting no operation. The corresponding product
operators of the 3 states are

\begin{equation}
\begin{array}{l}
{}\rho _1=I_{1z}+I_{2z}+I_{3z} \\ 
\rho _2=4I_{1z}I_{2z}I_{3z}+2I_{2z}I_{3z}-I_{3z} \\ 
\rho _3=2I_{1z}I_{3z}+2I_{1z}I_{2z}+I_{3z}
\end{array}
\end{equation}
The sum of the 3 states is

\begin{equation}
\rho _{sum}=\rho _{1}+\rho _{2}+\rho
_{3}=I_{1z}+I_{2z}+I_{3z}+2I_{1z}I_{3z}+2I_{1z}I_{2z}+2I_{2z}I_{3z}+4I_{1z}I_{2z}I_{3z}
\end{equation}
It can be seen from Eq. (7) and Eq. (10) that $\rho _{sum}$ is the same as $%
\rho _{eff}$ for $n=3$. The experimental pulse sequences was shown in Fig.
1. Of course, the same result to exclude this $1/2J_{13}$ evolution can also
be gained by swapping gates proposed by Collins et al.\cite{Collins}, but it
is more complex to implement experimentally. Note that a magnetic field
gradient pulses were accomplished at the end of the experiments in order to
eliminate the residual transverse magnetic vectors and thereby reduce the
errors.

The pseudo-pure state $\rho _{000}$ was achieved in the experiment, by
summing the three experimental results, where the qubits from high to low
corresponds to spin 1, 2 and 3, respectively. The resultant spectra of the
three spectra then recorded (shown in Fig.2) through reading-out pulses
confirm that a pseudo-pure state $\rho _{000}$ has been prepared. The
normalized diagonal elements of the deviation matrix $\rho _{000}$ for the
pseudo-pure state $|000\rangle $ by quantum state tomography \cite{Chuang4}
was given as

\begin{equation}
diag[\rho
_{000}]=[1.000,0.0314,-0.0291,-0.0032,0.0520,0.0114,-0.0535,-0.0277]
\end{equation}
All diagonal elements except one with the value of 1 in Eq. (11) should be
zero theoretically. So the maximal relative error of the experimental values
of the diagonal elements was shown to be $<6\%,$ with the small off-diagonal
elements$.$

Similarly, the procedure can be applied to a homonuclear 4-spin system $($i.
e. $n=4.)$ To obtain the pseudo-pure state $\rho _{0000},$ we can perform 5
experiments as follows

\begin{equation}
\begin{array}{l}
CN_{12}CN_{14}CN_{31},\qquad CN_{21}CN_{42}CN_{34},\qquad CN_{12}CN_{42} \\ 
CN_{12}CN_{14}N_{3},\qquad CN_{23}CN_{24}N_{31}
\end{array}
\end{equation}
Among a total of 20 terms produced, the 4 extra terms can be eliminated
pairwise through $N_{i}$ operations and one extra term removed through the
magnetic field gradient technique. For a homonuclear 5-spin system, the
procedure has clearly been demonstrated in Ref. \cite{Lieven}.

\subsection{The NMR results of the highly structure algorithm}

Applying the pulse sequences in Table. I. to the pseudo-pure state $\rho
_{000}$, we got the results of Hogg's algorithm. For $m=3$, because only the
diagonal elements are non-zeros, we reconstructed the normalized diagonal
elements like the $\rho _{000}$ preparation above. The experimental
deviation matrixes were obtained respectively 
\begin{equation}
\begin{array}{l}
diag[\rho _{V_{1}\wedge V_{2}\wedge
V_{3}}]=[-0.0190,0.0297,-0.0582,0.0631,-0.0072,0.0416,-0.0800,1.0000], \\ 
diag[\rho _{\bar{V}_{1}\wedge V_{2}\wedge
V_{3}}]=[0.0087,-0.0074,0.0959,-0.0845,0.0105,-0.0056,1.0000,-0.0393], \\ 
diag[\rho _{V_{1}\wedge \bar{V}_{2}\wedge
V_{3}}]=[-0.0412,0.0716,-0.0149,0.0187,-0.0580,1.0000,-0.0047,0.0289], \\ 
diag[\rho _{V_{1}\wedge V_{2}\wedge \bar{V}%
_{3}}]=[0.0037,0.0340,-0.0186,1.0000,-0.0211,0.0092,-0.0670,0.0881], \\ 
diag[\rho _{\bar{V}_{1}\wedge \bar{V}_{2}\wedge
V_{3}}]=[0.0512,0.0077,-0.157,-0.0269,1.0000,-0.0127,0.0029,-0.0082], \\ 
diag[\rho _{\bar{V}_{1}\wedge V_{2}\wedge \bar{V}%
_{3}}]=[0.0173,0.0171,1.0000,-0.0348,-0.0091,-0.0092,0.0606,-0.0093], \\ 
diag[\rho _{V_{1}\wedge \bar{V}_{2}\wedge \bar{V}%
_{3}}]=[-0.0304,1.0000,-0.0197,0.0200,-0.0228,0.0594,-0.0199,0.0199], \\ 
diag[\rho _{\bar{V}_{1}\wedge \bar{V}_{2}\wedge \bar{V}%
_{3}}]=[1.0000,0.0314,-0.0291,-0.0032,0.0520,0.0114,-0.0535,-0.0277].
\end{array}
\end{equation}

In contrast with the theoretical expectation, the maximal relative errors of
the experimental values of the diagonal elements was shown to be $<9\%,$
with the small off-diagonal elements$.$

For $m=1$, we reconstructed the experimental final deviation matrices by
quantum state tomography . In Fig. 4 were shown the theoretical and measured
results for these cases with the maximal relative errors $16\%$ $\sim 25\%$
of the density matrix elements. It can be seen clearly from Eq. (13) and
Fig. 4 that the experimental results are in good agreement with the theory.
Errors are primarily due to inhomogeneity of RF fields and static magnetic
fields, magnetization decay during the measurement and imperfect calibration
of the rotations.

\section{Discussion}

In summary, we have experimentally demonstrated a NMR realization of quantum
algorithm for a highly structured searching problem on a three-qubit quantum
computer. The simplified temporal averaging for preparing the pseudo-pure
states takes less separate experiments than the original temporal averaging
proposed by Knill et al.\cite{Knill}, e. g., 3 times instead of 7 times for
a 3-spin system. Moreover, its Signal-to-Noise ratio is better than that of
spatial averaging\cite{Cory1,Cory2}. Hogg's algorithm for 3 qubits was
completed on the prepared pseudo-pure state $\rho _{000}$, unlike some
algorithms only on the thermal state, and the final results were also read
out by a weak measurement on the ensemble, which accomplish the whole
process of quantum computation.

Hogg's algorithm based on the structure of a problem, with a particular
choice of the phases determined by the number of conflicts in assignments,
finds the solutions to 1-SAT problem in a single step with probability $%
100\% $ even for $n\rightarrow \infty $ . By contrast, the unstructured
search methods require $O(2^{n})$ steps classically, and $O(2^{n/2})$ steps
on quantum computers\cite{Grover}. For example, in the 3-qubit Grover's
algorithm NMR experiment implemented by Vandersypen et al.\cite{L.M.K} with
a heteronuclear system, per Grover iteration applied 19 pulses, 2 evolutions
of $1/2J$ and 3 of $1/4J$ and performed 2 iterations at least to find a
single solution $|x_{0}\rangle .$ Comparing our pulse sequences with it,
Hogg's algorithm adopts far less RF pulses and requires far less coherence
time so as to be realized more easily. In addition, probability $100\%$
guarantees that Hogg's algorithm is a complete method, i. e. failure to find
a solution definitely indicates the problem is not soluble. Unlike the
previously proposed quantum algorithms, such as the Grover's algorithm\cite
{Grover}, that find solutions with probability less than one, cannot
guarantee no solutions existing. Hogg's algorithm is more efficient to solve
any 1-SAT problem. Furthermore, this algorithm also applied to maximally
constrained soluble $k$-SAT problems\cite{Hogg1} for any $k$ with an
analogue procedure. To experimentally implement this algorithm for larger
systems, the main difficulties are to address and control the qubits well
and to maintain coherence during the computational process.

\begin{center}
{\bf ACKNOWLEDGEMENTS}
\end{center}

We thank Xiaodong Yang, Hanzheng Yuan, Xu Zhang and Guang Lu for help in the
course of experiments.

\begin{center}
{\bf Captions of the figures}
\end{center}

Fig. 1~~ Quantum circuit for 3-qubits that implements a highly structure
search algorithm. Horizontal lines represent qubits, time goes from left to
right. Using three Hadamard gates, an pseudo-pure states $|\psi _{0}\rangle
=|000\rangle $ is transformed into a uniform superposition state $|\psi
_{i}\rangle $, which is then converted to the answer state $|\psi _{f}>$
after the action of gates $R$ and $U$. For the definition of $R$ and $U$,
see text.

Fig. 2 NMR pulse sequences to implement the pseudo-pure state $\rho _{000}$.
Narrow and wide boxes correspond to $\frac{\pi }{2}$ and $\pi $ pulses
(refocusing pulses) respectively. $X$ and $Y$ denote the pulses along the x-
and y-axis, $\overline{X}$ and$\;\overline{Y},$ opposite to the x- and
y-axis. (a) to prepare the state of $\rho _{2}$ in Eq. (4). (b) to prepare
the state of $\rho _{3}$ in Eq. (4). $\rho _{1}$ is the thermal equilibrated
state, no operation). The pulse sequences are designed for alanine with
small $J_{13}.$

Fig. 3\quad The resultant $^{13}C$ spectra of three experimentally measure
spectra of the prepared pseudo-pure state $\rho _{000}.$ The reading-out
pulses (a) $\left[ \frac{\pi }{2}\right] _{y}^{1}$, (b) $\left[ \frac{\pi }{2%
}\right] _{y}^{2},$ (c) $\left[ \frac{\pi }{2}\right] _{y}^{3}$ were applied
respectively. The abscissa indicates the frequency, and the ordinate denotes
the intensity of the spectra (in arbitrary unit).

Fig. 4~~ Experimental and theoretical deviation density matrices (in
arbitrary units) for the Hogg's algorithm of all logic formulas when $m=1$.
(a)---(f) represent the recovered matrices for the logic formula $V_{1},\bar{%
V}_{1},V_{2},\bar{V}_{2},V_{3},\bar{V}_{3},$respectively, the left and right
column denoting the real and imaginary components. (a1)---(f1) are the
corresponding theoretical values. Relative errors are shown as
percentages.\newpage

\begin{center}
{\bf Captions of the tables}
\end{center}

Table. I All the logic formulas for 1-SAT of a 3-spin system for $m=1$ $or$ $%
3$ , the corresponding theoretical solutions and the reduced pulse sequences
for the operator $URW,$ where the logic variables with subscript $i$ stand
for spin $i.$ $X$ and $Y$ denote the $\frac{\pi }{2}$ pulses along the x-
and y-axis, $\overline{X}$ and$\;\overline{Y},$ opposite to the x- and
y-axis. The subscripts represent the qubits.

\begin{tabular}{|l|c|c|l|}
\hline
$m$ & the logic formula & the corresponding & the corresponding reduced \\ 
&  & \multicolumn{1}{|c|}{theoretical solutions} & pulses sequence for $URW$
\\ \hline
& \multicolumn{1}{|l|}{$V_{1}$} & \multicolumn{1}{|l|}{$|001\rangle
+|011\rangle +|101\rangle +|111\rangle $} & $X_{1}^{2}Y_{2}Y_{3}$ \\ 
\cline{2-4}
& \multicolumn{1}{|l|}{$\bar{V}_{1}$} & \multicolumn{1}{|l|}{$|000\rangle
+|010\rangle +|100\rangle +|110\rangle $} & $Y_{2}Y_{3}$ \\ \cline{2-4}
$1$ & \multicolumn{1}{|l|}{$V_{2}$} & \multicolumn{1}{|l|}{$|010\rangle
+|011\rangle +|110\rangle +|111\rangle $} & $Y_{1}X_{2}^{2}Y_{3}$ \\ 
\cline{2-4}
& \multicolumn{1}{|l|}{$\bar{V}_{2}$} & \multicolumn{1}{|l|}{$|000\rangle
+|001\rangle +|100\rangle +|101\rangle $} & $Y_{1}Y_{3}$ \\ \cline{2-4}
& \multicolumn{1}{|l|}{$V_{3}$} & \multicolumn{1}{|l|}{$|100\rangle
+|101\rangle +|110\rangle +|111\rangle $} & $Y_{1}Y_{2}X_{3}^{2}$ \\ 
\cline{2-4}
& \multicolumn{1}{|l|}{$\bar{V}_{3}$} & \multicolumn{1}{|l|}{$|000\rangle
+|001\rangle +|010\rangle +|011\rangle $} & $Y_{1}Y_{2}$ \\ \hline
& \multicolumn{1}{|l|}{$V_{1}\wedge V_{2}\wedge V_{3}$} & 
\multicolumn{1}{|l|}{$|111\rangle $} & $(X\bar{Y}X)_{1}(X\bar{Y}X)_{2}(X\bar{%
Y}X)_{3}$ \\ \cline{2-4}
& \multicolumn{1}{|l|}{$\bar{V}_{1}\wedge V_{2}\wedge V_{3}$} & 
\multicolumn{1}{|l|}{$|110\rangle $} & $(X\bar{Y}\bar{X})_{1}(X\bar{Y}%
X)_{2}(X\bar{Y}X)_{3}$ \\ \cline{2-4}
& \multicolumn{1}{|l|}{$V_{1}\wedge \bar{V}_{2}\wedge V_{3}$} & 
\multicolumn{1}{|l|}{$|101\rangle $} & $(X\bar{Y}X)_{1}(X\bar{Y}\bar{X}%
)_{2}(X\bar{Y}X)_{3}$ \\ \cline{2-4}
$3$ & \multicolumn{1}{|l|}{$\bar{V}_{1}\wedge \bar{V}_{2}\wedge V_{3}$} & 
\multicolumn{1}{|l|}{$|100\rangle $} & $(X\bar{Y}\bar{X})_{1}(X\bar{Y}\bar{X}%
)_{2}(X\bar{Y}X)_{3}$ \\ \cline{2-4}
& \multicolumn{1}{|l|}{$V_{1}\wedge V_{2}\wedge \bar{V}_{3}$} & 
\multicolumn{1}{|l|}{$|011\rangle $} & $(X\bar{Y}X)_{1}(X\bar{Y}X)_{2}(X\bar{%
Y}\bar{X})_{3}$ \\ \cline{2-4}
& \multicolumn{1}{|l|}{$\bar{V}_{1}\wedge V_{2}\wedge \bar{V}_{3}$} & 
\multicolumn{1}{|l|}{$|010\rangle $} & $(X\bar{Y}\bar{X})_{1}(X\bar{Y}%
X)_{2}(X\bar{Y}\bar{X})_{3}$ \\ \cline{2-4}
& \multicolumn{1}{|l|}{$V_{1}\wedge \bar{V}_{2}\wedge \bar{V}_{3}$} & 
\multicolumn{1}{|l|}{$|001\rangle $} & $(X\bar{Y}X)_{1}(X\bar{Y}\bar{X}%
)_{2}(X\bar{Y}\bar{X})_{3}$ \\ \cline{2-4}
& \multicolumn{1}{|l|}{$\bar{V}_{1}\wedge \bar{V}_{2}\wedge \bar{V}_{3}$} & 
\multicolumn{1}{|l|}{$|000\rangle $} & $(X\bar{Y}\bar{X})_{1}(X\bar{Y}\bar{X}%
)_{2}(X\bar{Y}\bar{X})_{3}$ \\ \hline
\end{tabular}


\begin{references}
\bibitem{Deutsch}  D. Deutsh, and R. Jozsa, Proc. Roy. Soc. Lond. A, 439,
553 (1992).

\bibitem{Cleve}  R. Cleve, A. Ekert, C. Macchiavello and M. Mosca, Proc.
Roy. Soc. Lond. A, 454, 339 (1998).

\bibitem{Shor}  P.Shor, Algorithms for quantum computation: discrete
logarithms and factoring. proc. 35th Annu. Symp. on Found. of Computer
Science, (IEEE comp. Soc. Press, Los Alomitos,CA. 1994) 124-134.

\bibitem{Grover}  L. K. Grover, Phys. Rev. Lett. 79, 325 (1997).

\bibitem{Cirac}  J. Cirac and P. Zoller, Phys. Rev. Lett. 74, 4091 (1995).

\bibitem{Monroe}  C. Monroe, D. M. Meekhof, B. E. King, W. M. Itano, and D.
J. Wineland, Phys. Rev. Lett. 75, 4714 (1995).

\bibitem{Bandyopadhyay}  S. Bandyopadhyay and V. Roychowdhury, Jpn. J. Appl.
Phys. 35, 3550 (1996).

\bibitem{Domokos}  P. Domokoss, J. M. Raimond, M. Brune, S. Haroche, Phys.
Rev. Lett. 52, 3554 (1995).

\bibitem{Turchette}  Q. A. Turchette, C. J. Hood, W. Lange, H. Mabuchi, and
H. J. Kimble, Phys. Rev. Lett, 75, 4710 (1995).

\bibitem{Cory1}  D. G. Cory, A. F. Fahmy and T. F. Havel, Proc. Natl. Acad,
Sci. USA 94 (1997) 1634 .

\bibitem{Gershenfeld}  N. Gershenfeld and I. L. Chuang, Science 275, 350
(1997).

\bibitem{Cory2}  D.G. Cory, M. D. Price and T. F. Havel, Physica D 120
(1998) 82.

\bibitem{Knill}  E. Knill, I.Chuang and R. Laflamme, Phys. Rev. A 57 (1998)
3348.

\bibitem{ChuangDJ}  I. L. Chuang, L. M. K. Vandersypen, X. Zhou, D. W.
Leung, S. Lloyd, Nature 393, 143 (1998).

\bibitem{Jones1}  J. A. Jones and Mosca, J. Chem. phys. 109, 1648 (1998).

\bibitem{Linden}  N. Linden, H. Barjat and R. Freeman, Chem. Phys. Lett. 80,
3408, 229 (1998).

\bibitem{Kim}  J. Kim, J.-S. Lee and S. Lee, Phys. Rev. A 62, 022312 (2000).

\bibitem{Collins}  D. Collins, K. W. Kim, W. C. Holton, H.
sierzputowska-Gracz, and E. O. Stejskal, Phys. Rev. A 62, (2000).

\bibitem{Marx}  R. Marx, A. F. Fahmy, J. M. Myers, W. Bermel and s. J.
Glaser, Phys. Rev. A 62, 012310 (2000).

\bibitem{Chuang3}  I. L. Chuang, N. Gershenfeld and M. Kubinec, Phys. Rev.
Lett. 80, 3408 (1998).

\bibitem{Jones2}  J. A. Jones, m. Mosca and R. H. Hansen, Nature 393. 344
(1998).

\bibitem{L.M.K}  L. M. K. Vandersypen et al., Appl. Phys. Lett. 76, 646
(2000).

\bibitem{Knill1}  E. Knill, R. Laflamme, R. Martinez and C.-H. tsieng,
Nature 404, 368 (2000).

\bibitem{Lieven}  Lieven M. K, Vandersypen, Matthias Steffen, and Gregory
Breyta, Phys. Rev. Lett. 85, 5452 (2000).

\bibitem{Hogg1}  T.Hogg, Phys. Rev. Lett. 80 (1998) 2473.

\bibitem{Zhu}  X. Zhu, X. Fang, M. Feng, F. Du, K. Gao, and X. Mao, Physica
D 156, 179 (2001).

\bibitem{Garey}  M. R. Garey and D. S. Johnson, Computers and
Intractability: a Guide to the Theory of NP-Completeness, Freeman, San
Francisco, (1979).

\bibitem{Ernst}  R. Ernst, G. Bodenhausen and A. Wokaun, Principles of
Nuclear Magnetic Resonance in One and Two Dimensions (Oxford Univ. Press,
Oxford, 1990).

\bibitem{Dorai}  K. Dorai, Arvind and A. Kumar, Phys. Rev. A 61 (2000)
042306.

\bibitem{Chuang4}  I. L. Chuang, N. Gershenfeld, M. Kubinec and D. Leung,
Proc. Roy. Soc.Lond A 454 (1998) 447.

\bibitem{Chuang1}  I. L. Chuang, and Y. Yamamoto, Phys. Rev. A 52, 3489
(1995).

\bibitem{Jones3}  J. A. Jones, Science 280, 229 (1998).\newpage
\end{references}
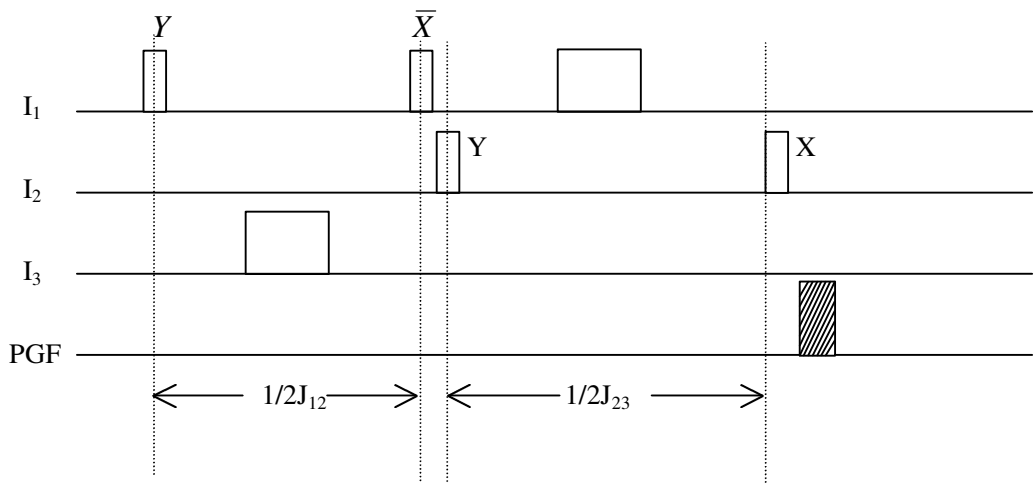
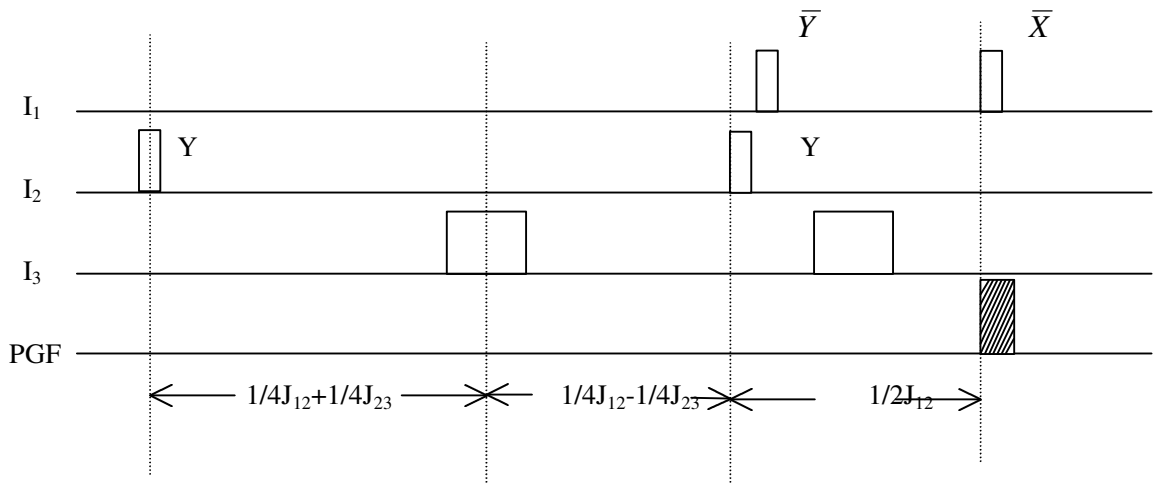
\end{document}